\documentclass[11pt]{article}
\usepackage{color}
\usepackage{subfigure}
\usepackage{amsmath}
\usepackage{amssymb}
\usepackage{graphicx,color}
\usepackage{cite}
\usepackage{enumerate}
\usepackage{amsthm}
\usepackage{amsfonts,mathrsfs}
\usepackage{geometry}
\usepackage{ifthen}
\usepackage{graphicx}
\usepackage{epstopdf}
\usepackage{float}
\usepackage{bm}
\usepackage[caption=false]{subfig}
\usepackage[utf8]{inputenc}

\begin{document}

\title{\bf Imaginarity as a resource in Shor's algorithm}

\vskip0.1in
\author{\small Linlin Ye$^1$, Chuanfa Wu$^1$, Zhaoqi Wu$^1$\thanks{Corresponding author. E-mail: wuzhaoqi\_conquer@163.com},
Shao-Ming Fei$^{2,3}$\\
{\small\it  1. Department of Mathematics, Nanchang University,
Nanchang 330031, P R China}\\
{\small\it  2. School of Mathematical Sciences, Capital Normal University, Beijing 100048, P R China}\\
{\small\it  3. Max-Planck-Institute for Mathematics in the Sciences,
04103 Leipzig, Germany} }

\date{}
\maketitle

\noindent {\bf Abstract} {\small }\\
We investigate the sequential version of Shor's order-finding
algorithm from the perspective of the imaginarity of quantum
channels. By fixing an eigenbranch of the modular multiplication
unitary and a target output path, the process is effectively
decomposed into a real diagonal factor and a residual phase channel.
We find that for the ideal residual phase channel, the three
trace-norm induced imaginarity measures are all equal. On good
output paths, a smaller residual imaginarity corresponds to a higher
single-block success probability. By combining the branch-wise
probabilities, we obtain the exact ideal success probability of the
sequential order-finding protocol and derive lower and upper bounds
in terms of residual phase-channel imaginarity. We further extend
the analysis to noisy residual phase channels, and obtain the lower
and upper bounds on the noisy success probability. These results
clarify the role of imaginarity of channels in Shor's algorithm and
provide a refined perspective on quantum-resource behavior in ideal
and noisy sequential quantum algorithms.

 \vskip0.2in

\noindent {\bf 1. Introduction}\\\hspace*{\fill}\\
Ever since the inception of quantum mechanics, the fundamental necessity of complex numbers has been a subject of enduring debate. Given that observable physical phenomena are ultimately described by real-valued probabilities, early researchers argued that complex numbers are not theoretically indispensable in quantum mechanics, and merely serve to simplify calculations \cite{Birkhoff1936QuantumLogic,Stueckelberg1960RealQM,McKague2009RealHilbert,Dyson1962ThreefoldWay,
Aleksandrova2013RealVector}. Renou et al.~\cite{Renou2021RealNumbers} tested an entanglement-swapping scenario and found that complex-valued quantum mechanics agrees with the experimental data, whereas real-valued quantum mechanics exhibits a clear deviation. These results provide experimental evidence for the indispensable role of complex numbers in quantum mechanics. Subsequent experiments offered direct refutations of  ``real-only'' quantum mechanics, including a superconducting-qubit implementation that significantly violated the bounds implied by real quantum theory \cite{Chen2022RealValuedQM}, as well as a photonic network test performed under strict assumptions of locality and independent sources, which also ruled out real-valued quantum mechanics \cite{Wu2022RealValuedQM}.

To explore the significance of complex numbers in quantum systems and their potential advantages, Hickey and Gour \cite{HickeyGour2018Imaginarity} proposed to treat the imaginarity of quantum states as an independent quantum resource, thereby initiating the framework of the resource theory of imaginarity. Wu et al. \cite{Wu2021OperationalImaginarity,Wu2021ResourceImaginarity} demonstrated theoretically and experimentally that complex numbers play an essential role in quantum-state discrimination. Related developments have further deepened our understanding of imaginarity, including unitary-invariant witnesses of quantum imaginarity \cite{Fernandes2024UnitaryWitnesses} and multistate imaginarity in qubit systems \cite{Li2026MultistateImaginarity}. Beyond its foundational significance, imaginarity has also been shown to be relevant to a variety of quantum-information tasks and physical scenarios, such as quantum machine learning \cite{Sajjan2023ImaginaryComponents}, pseudorandom unitary operators \cite{Haug2025PseudorandomUnitaries}, quantum speed limits \cite{Xuan2025QuantumImaginarityQSL} and circuit complexity \cite{Ye2026CircuitComplexity}.
Parallel to these applications, considerable effort has been devoted to quantifying the imaginarity of quantum states from different physical perspectives. Representative measures include the robustness of imaginarity \cite{HickeyGour2018Imaginarity,Wu2021ResourceImaginarity}, the $l_1$ norm of imaginarity \cite{Chen2023ImaginarityMeasures}, geometric-like imaginarity \cite{Guo2025GeometricLikeImaginarity} and distance-based imaginarity measures \cite{GuoHuangLiFei2025AQT}. More recently, the notion of imaginarity has also been extended beyond quantum states to quantum operations, leading to quantitative studies of the imaginarity of quantum operations \cite{Wu2026QuantumOperations}. Beyond the construction of imaginarity measures, another important line of research concerns the convertibility of imaginary states under free operations, which clarifies the operational structure of imaginarity as a resource \cite{HickeyGour2018Imaginarity,Wu2021ResourceImaginarity,Guo2025GeometricLikeImaginarity,Du2025PureStateImaginarity}. In a further development, Zhang et al. \cite{Zhang2023ImaginaringChannels} extended the discussion from states to quantum channels by introducing and analyzing the capacities of qubit channels in generating or eliminating imaginarity.

Quantum algorithms are designed to exploit quantum-mechanical principles to enhance the efficiency of solving certain computational problems on quantum computers \cite{NielsenChuang2000QuantumComputation}. Representative examples include the Deutsch--Jozsa algorithm for determining whether a Boolean function is constant or balanced \cite{DeutschJozsa1992RapidSolution,CollinsKimHolton1998DeutschJozsa}, Simon's algorithm for solving quantum oracle problems \cite{Simon1997PowerQuantumComputation,TameBellDiFrancoWadsworthRarity2014Simon,GhoshSarkar2021Simon,YeWuFei2023Simon}, Grover's algorithm for unstructured search \cite{Grover1997NeedleHaystack,YeWuFei2023Grover}, the Bernstein--Vazirani algorithm for identifying a hidden bit string \cite{DuShiZhouFanYeHanWu2001BernsteinVazirani}, and the HHL algorithm for solving systems of linear equations \cite{HarrowHassidimLloyd2009HHL,YeWuFei2023HHL}.
Among these algorithms, Shor's algorithm \cite{Shor1997Factorization} occupies a central position as it provides an efficient quantum approach to integer factorization and achieves a superpolynomial speedup over the best known classical algorithms. The factorization problem can be reduced to order finding: given an integer $N$ to be factored and an integer $x$ coprime to $N$, one seeks the smallest positive integer $r$ satisfying
$x^r \equiv 1 \pmod N,$ or equivalently, the period of the function $x^a \bmod N$. Since the security of RSA-type public-key cryptosystems relies on the presumed classical hardness of integer factorization, Shor's algorithm has important implications for cryptographic security.

This potential impact has stimulated extensive research on more resource-efficient implementations of Shor's algorithm  \cite{FowlerHollenberg2004Rotation,NamBlumel2013BandedQFT,
MartinLopezLaingLawson2012QubitRecycling,WengChuu2024Shor}, such as the qubit-efficient circuit constructions for Shor's algorithm \cite{Beauregard2003Shor,HanerRoettelerSvore2017Shor} and alternative resource models for efficient factorization using one pure qubit assisted by mixed qubits \cite{ParkerPlenio2000}. Further studies have also examined how Shor's algorithm could be implemented in realistic quantum computing models to attack the RSA encryption \cite{KuteDesaiJadhav2024RSA}. Nevertheless, current resource estimates indicate that factoring a standard 2048-bit RSA modulus would still require fault-tolerant quantum hardware with approximately millions of physical qubits \cite{GidneyEkera2021RSA,Gidney2025RSA}. At the experimental level, Monz et al. \cite{MonzNiggMartinez2016Shor} implemented a scalable version of Shor's algorithm using a small trapped-ion quantum computer composed of five calcium ions, providing useful guidance for resource-conscious quantum circuit design.
In addition to implementation and resource analysis, the quantum resources involved in Shor's algorithm have also attracted increasing attention. In particular, the coherence, decoherence and entanglement-related dynamics of Shor's algorithm have been investigated \cite{YeWuZhou2026Shor,YeWuFei2026CTP}. Moreover, Ahnefeld et al.\cite{Ahnefeld2022Shor} studied a sequential version of Shor's algorithm that preserves a fixed overall structure and used channel coherence to derive lower and upper bounds on the success probability of the protocol.

The remainder of this paper is organized as follows. In Section 2, we review Shor's algorithm and recall imaginarity measures of quantum operations based on the trace norm.
In Section 3, we derive the branch-wise residual phase left after the adaptive feed-forward correction and obtain the corresponding
single-block success probability. In Section 4, we compute the imaginarity of the residual phase channel and derive the bound on full ideal
order-finding success probability. In Section 5, we extend the analysis to noisy residual phase channels, and present the corresponding bounds for the noisy success probability.
We summarize the results in Section 6.

\vskip0.2in
\noindent {\bf 2. Shor's algorithm and imaginarity of quantum channels}\\

\noindent{\bf 2.1 Shor's algorithm}

We first briefly review the order-finding part of Shor's algorithm.
Let $N$ be the odd composite integer to be factored, and $x<N$ an integer satisfying $\gcd(x,N)=1$. The order-finding problem aims to determine the smallest positive integer $r$ such that
$x^r \equiv 1 \pmod N .$ The order $r$ is recovered from $K/q$ by the continued-fraction algorithm. If $r$ is even and $x^{r/2}\not\equiv -1 \pmod N$, the factors of $N$ are then obtained from $\gcd(x^{r/2}\pm1,N)$. The quantum system consists of two registers: the control register $A$ and the work register $B$. The control register $A$ contains $L$ qubits, and we denote
$q=2^L .$ The work register $B$ is used to store the results of the modular multiplication. The initial state of the whole system is chosen as
$|0\rangle_A^{\otimes L}|1\rangle_B .$ That is, the control register is initially prepared in the all-zero state, while the work register is initialized in $|1\rangle_B$ \cite{Shor1997Factorization}.

First, Hadamard gates are applied to all $L$ qubits in the control register $A$. This operation transforms the all-zero state into a uniform superposition:
$
H^{\otimes L}|0\rangle^{\otimes L}
=
\frac{1}{\sqrt q}\sum_{m=0}^{q-1}|m\rangle .
$
Consequently, after applying the Hadamard gates, the state of the entire system becomes
\begin{equation}
|\psi_1\rangle
=
\frac{1}{\sqrt q}
\sum_{m=0}^{q-1}
|m\rangle_A |1\rangle_B .
\end{equation}
This step prepares the control register in a superposition over all possible exponents $m$, which enables the subsequent modular exponentiation to be performed coherently.

Next, the modular multiplication unitary on the work register $B$ is defined by
$U_B|y\rangle = |xy \bmod N\rangle .$
The states $|1\rangle, |x\rangle, \ldots, |x^{r-1}\rangle$ span a cyclic subspace. Since $x^r \equiv 1 \pmod N$, one has $U_B^r = I$ on this cyclic subspace.
The global controlled modular multiplication operation in Shor's order-finding algorithm is defined as
$U=
\sum_{m=0}^{q-1}
|m\rangle\langle m|_A\otimes U_B^m .$
This operation means that, conditioned on the control register being in the computational basis state $|m\rangle_A$, the unitary $U_B^m$ is applied to the work register. Since
$U_B^m|1\rangle_B
=|x^m \bmod N\rangle_B$, after the controlled modular multiplication the state of the system becomes
\begin{equation}
|\psi_2\rangle
=
\frac{1}{\sqrt q}
\sum_{m=0}^{q-1}
|m\rangle_A |x^m \bmod N\rangle_B .
\end{equation}
This step correlates the exponent $m$ in the control register with the function value $x^m \bmod N$ in the work register, thereby encoding the period $r$ into the quantum state.

To analyze the phase structure of this state, we introduce the eigenstates of $U_B$ on the cyclic subspace. For $j=0,1,\ldots,r-1$, define
\begin{equation}
|\psi_j\rangle_B
=
\frac{1}{\sqrt r}
\sum_{a=0}^{r-1}
e^{-2\pi \mathrm{i}ja/r}
|x^a \bmod N\rangle_B .
\end{equation}
It is straightforward to verify that
$U_B|\psi_j\rangle_B
=
e^{2\pi \mathrm{i}j/r}
|\psi_j\rangle_B .$
Thus, $|\psi_j\rangle_B$ is an eigenstate of $U_B$ with eigenvalue $e^{2\pi \mathrm{i}j/r}$. Moreover, the initial work-register state can be decomposed as
$|1\rangle_B
=
\frac{1}{\sqrt r}
\sum_{j=0}^{r-1}
|\psi_j\rangle_B .$
This eigenbasis representation provides a phase-estimation interpretation of the controlled modular multiplication. On a fixed eigenbranch $j$, one has
$U_B^m|\psi_j\rangle_B
=
e^{2\pi \mathrm{i}mj/r}
|\psi_j\rangle_B .$
Therefore, the action of the global controlled modular multiplication on this eigenbranch is given by
\begin{equation}
|m\rangle_A|\psi_j\rangle_B
\longmapsto
e^{2\pi \mathrm{i}mj/r}|m\rangle_A|\psi_j\rangle_B .
\end{equation}
This shows that, on each eigenbranch $j$, the work register itself remains unchanged except for the eigenphase contribution, while the phase factor $e^{2\pi \mathrm{i}mj/r}$ is imprinted on the control register. In this sense, Shor's algorithm estimates the phase $j/r$ by means of quantum phase estimation and then recovers the order $r$ through classical continued-fraction post-processing.
Substituting the eigenstate decomposition of the initial work-register state into $|\psi_2\rangle$, the post-modular-multiplication state can be rewritten as
\begin{equation}
|\psi_2\rangle
=
\frac{1}{\sqrt r}
\sum_{j=0}^{r-1}
\left(
\frac{1}{\sqrt q}
\sum_{m=0}^{q-1}
e^{2\pi \mathrm{i}mj/r}|m\rangle_A
\right)
|\psi_j\rangle_B .
\end{equation}
Hence, in the $j$-th eigenbranch, the control register carries phase information about the fraction $j/r$.

Finally, the inverse quantum Fourier transform $F_q^\dagger$ is applied to the control register $A$. The role of this operation is to convert the phase information stored in the amplitudes of the control register into a probability distribution over computational-basis measurement outcomes. After the inverse Fourier transform, measuring the control register yields an integer $K\in\{0,1,\ldots,q-1\}$ with high probability near values satisfying
\begin{equation}
\frac{K}{q}\approx \frac{j}{r}.
\end{equation}
Therefore, by measuring the control register and applying the classical continued-fraction algorithm to $K/q$, one can recover the fraction $j/r$ and hence obtain the order $r$. This completes the quantum order-finding part of Shor's algorithm.

The final measurement output of the control register is an integer
$K\in\{0,1,\ldots,q-1\}
=
\{0,1,\ldots,2^L-1\}.$
We use the least-significant-bit convention,
\begin{equation}
K=\sum_{t=0}^{L-1}2^t k_t,
~~
k_t\in\{0,1\}.
\end{equation}
Thus $k_0$ is the least significant bit and $k_{L-1}$ is the most significant bit.

In the sequential version of the order-finding algorithm \cite{Ahnefeld2022Shor,ParkerPlenio2000}, the inverse Fourier transform is implemented bit by bit. The $l$-th block uses the modular unitary,
$U_l=U_B^{2^{L-l}}$, $l=1,2,\ldots,L$. The corresponding circuit structure is illustrated in Fig.~1.
With the above convention for the binary expansion of a target output $K$, the bit associated with the $l$-th block is $k_{l-1}$. In an actual run of the sequential circuit, the measurement outcome of the $l$-th block is denoted by $b_{l-1}$. Along the target path $K$, successful readout of the $l$-th block means $b_{l-1}=k_{l-1}$.
\begin{figure}[ht]
\centering {\begin{minipage}[figure1]{1.0\linewidth}
\includegraphics[width=1.0\textwidth]{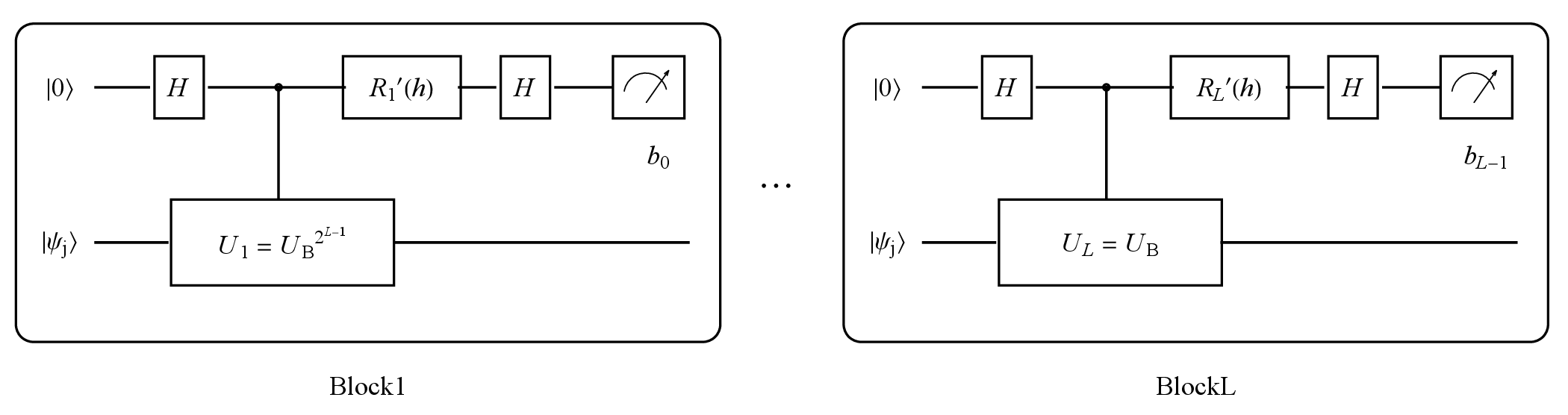}
\end{minipage}}
\caption{
Sequential circuit for the order-finding part of Shor's algorithm.
The \(l\)-th block uses the controlled unitary \(U_l=U_B^{2^{L-l}}\), an adaptive
phase rotation \(R'_l(h)\), and a final measurement producing the bit \(b_{l-1}\),
while the work register is analyzed on a fixed eigenbranch \(|\psi_j\rangle\).
}
\end{figure}

\noindent{\bf 2.2 Imaginarity of quantum channels}

In the resource theory of imaginarity, a reference orthonormal basis
is fixed in advance.  A quantum state is called a real state, or a free
state, if all the entries of the density matrix are real with respect to the fixed
basis. Equivalently, a state $\rho$ is free if it satisfies
$\rho=\rho^T$, where the transposition $T$ is taken in the fixed
reference basis. States containing nonzero imaginary matrix elements
are therefore regarded as resourceful states. To extract the real
part of a quantum state, one introduces the deimaginarity map,
\begin{equation}
\Delta(\rho)=\frac{\rho+\rho^T}{2}=\sum_{i,j}\mathrm{Re}(\rho_{ij})|i\rangle\langle j|
\end{equation}
for $\rho=\sum_{i,j}\rho_{ij}|i\rangle\langle j|$,
that is, $\Delta$ removes the imaginary parts of $\rho$ and keeps
only its real components.

Let $\mathcal{FC}$ denote the set of free channels in the resource
theory of imaginarity. For a quantum channel $\Theta$, the
trace-norm type dynamical imaginarity measures are defined
as \cite{Wu2026QuantumOperations},
\begin{equation}
M_c^t(\Theta) = \min_{\Phi\in \mathcal{FC}}\max_{\rho} \left\|
(\Theta\Delta-\Phi\Delta)(\rho) \right\|_1,
\end{equation}
\begin{equation}
M_d^t(\Theta) = \min_{\Phi\in \mathcal{FC}}\max_{\rho} \left\|
(\Delta\Theta-\Delta\Phi)(\rho) \right\|_1
\end{equation}
and
\begin{equation}
M_{dc}^t(\Theta)
=
\max_{\rho}
\left\|
(\Delta\Theta-\Theta\Delta)(\rho)
\right\|_1.
\end{equation}
Here $M_c^t(\Theta)$ quantifies the ability of $\Theta$ to create
imaginarity from real inputs, and $M_d^t(\Theta)$ characterizes the
ability of $\Theta$ to detect the imaginarity contained in input
states. In contrast, $M_{dc}^t(\Theta)$ measures the
noncommutativity between $\Theta$ and the deimaginarity map
$\Delta$, namely, the maximal trace-norm difference between applying
the channel before deimaginarization and applying deimaginarization
before the channel. If these quantities vanish, the channel is
indistinguishable from a free channel in the corresponding sense.
Larger values indicate stronger dynamical imaginarity of the channel.

\vskip0.2in

\noindent {\bf 3. Residual phase and single-block success probability}

In this section, we study the branch-wise action of a single block in the sequential order-finding protocol.
For a fixed eigenbranch $j$ and target output path $K$, we show that the feed-forward correction leaves a residual phase $\beta_{l,j,K}$ after cancelling the known tail phase.
This leads to an effective decomposition into a real diagonal factor and a residual phase channel.
We then derive the corresponding control-qubit state and the single-block success probability.

In the sequential version, the $l$-th block uses a single control qubit and applies the controlled unitary
$CU_l=|0\rangle\langle 0|\otimes I+|1\rangle\langle 1|\otimes U_l$,
where $U_l=U_B^{2^{L-l}}.$ Note that $CU_l$ is a joint unitary
operator on the control qubit and the work register. However, after
fixing an eigenbranch of the work register, it induces an effective
single-qubit phase gate on the control qubit. Define
\begin{equation*}
\theta_{l,j}
=
2\pi 2^{L-l}\frac{j}{r}.
\end{equation*}
Then, for a general control qubit state $a|0\rangle+b|1\rangle$, we have
\begin{equation}
CU_l
\left[
(a|0\rangle+b|1\rangle)\otimes|\psi_j\rangle_B
\right]
=
\left(
a|0\rangle
+
b e^{\mathrm{i}\theta_{l,j}}|1\rangle
\right)
\otimes|\psi_j\rangle_B.
\end{equation}
Consequently, on the fixed eigenbranch $j$, the controlled operation $CU_l$ induces the effective single-qubit phase operator
$V_{\theta_{l,j}}
=
\operatorname{diag}(1,e^{\mathrm{i}\theta_{l,j}})$
on the control qubit. Equivalently, for any control-qubit state $\rho$, the induced single-qubit channel is
$\mathcal U_{l,j}(\rho)=V_{\theta_{l,j}}\rho V_{\theta_{l,j}}^{\dagger}$.
This should be understood as a branch-wise effective description.
$CU_l$ is a joint unitary operator on the control qubit and the work register. However,
once the work register is restricted to the eigenstate $|\psi_j\rangle_B$, it contributes only the eigenphase
$e^{\mathrm{i}\theta_{l,j}}$, so that the action on this branch is
equivalently represented as a phase gate on the control qubit.

The adaptive feed-forward phase in the $l$-th block is determined by the previously measured bits. Let
$h=(b_0,\ldots,b_{l-2})$
denote the measurement history before the $l$-th block. If the history agrees with the lower bits of the target output path $K$, namely,
$h=(k_0,\ldots,k_{l-2})$, the feed-forward tail phase is
\begin{equation*}
\phi_l'(K)
=
\frac{k_{l-2}}{2^2}
+
\frac{k_{l-3}}{2^3}
+
\cdots
+
\frac{k_0}{2^l}.
\end{equation*}
For $l=1$, we set $\phi_1'(K)=0.$ The feed-forward phase gate is
$R_l'(K)=|0\rangle\langle 0|+e^{-2\pi \mathrm{i}\phi_l'(K)}|1\rangle\langle 1|$.
The corresponding single-qubit channel is
$\mathcal R_{l,h}(\rho)
=
R_l'(h)\rho R_l'(h)^{\dagger}.$
Define the residual phase
\begin{equation*}
\beta_{l,j,K}
=
2\pi 2^{L-l}
\left(
\frac{j}{r}
-
\frac{K}{2^L}
\right).
\end{equation*}

\noindent{\bf Theorem 1}
The induced phase $V_{\theta_{l,j}}$ and the feed-forward phase $R_l'(K)$ satisfy the following relation,
\begin{equation}
R_l'(K)V_{\theta_{l,j}}=Z^{k_{l-1}}V_{\beta_{l,j,K}},
\end{equation}
where $Z=\operatorname{diag}(1,-1), V_{\beta_{l,j,K}}=\operatorname{diag}(1,e^{\mathrm{i}\beta_{l,j,K}}).$

\noindent{\it Proof.}
From the binary expansion of $K$, we have
\begin{equation*}
\frac{K}{2^l}
=\sum_{t=0}^{L-1}2^{t-l}k_t=
M_l+\frac{k_{l-1}}{2}+\phi_l'(K),
\end{equation*}
where
\begin{equation*}
M_l=\sum_{t=l}^{L-1}2^{t-l}k_t\in\mathbb Z.
\end{equation*}
By using the following identity,
\begin{equation*}
2\pi 2^{L-l}\frac{j}{r}
=
2\pi 2^{L-l}\frac{K}{2^L}
+
2\pi 2^{L-l}
\left(
\frac{j}{r}
-
\frac{K}{2^L}
\right),
\end{equation*}
we have
\begin{equation*}
2\pi 2^{L-l}\frac{j}{r}
=
2\pi M_l
+
\pi k_{l-1}
+
2\pi\phi_l'(K)
+
\beta_{l,j,K}.
\end{equation*}
Therefore,
\begin{equation*}
R_l'(K)V_{\theta_{l,j}}
=\operatorname{diag}\left(1,e^{\mathrm{i}(2\pi M_l+\pi k_{l-1}+\beta_{l,j,K})}\right)
=
Z^{k_{l-1}}
V_{\beta_{l,j,K}},
\end{equation*}
where $Z=\operatorname{diag}(1,-1)$ and $V_{\beta_{l,j,K}}
=\operatorname{diag}(1,e^{\mathrm{i}\beta_{l,j,K}}).$ Note that since the
factor $Z^{k_{l-1}}$ is a real diagonal operation, it is a free
operation in the resource theory of imaginarity. \hfill $\square$

The corresponding channels are
$\mathcal Z_{k_{l-1}}(\rho)
=
Z^{k_{l-1}}\rho Z^{k_{l-1}}$
and
$\Theta_{\beta_{l,j,K}}(\rho)
=
V_{\beta_{l,j,K}}\rho V_{\beta_{l,j,K}}^{\dagger}.$
Thus, after fixing $j$, the correct history $h$ and the output path $K$, the middle part of the induced channel chain can be written as
$\mathcal R_{l,h}\circ\mathcal U_{l,j}=\mathcal Z_{k_{l-1}}\circ\Theta_{\beta_{l,j,K}}$.
This decomposition should be understood as an effective branch-wise representation, rather than as a modification of the physical circuit. Since $Z^{k_{l-1}}$ and $V_{\beta_{l,j,K}}$ are both diagonal operators, the corresponding channels commute. Therefore, in the branch-wise representation, the order of these two factors does not affect the result.
The channel
$\Theta_{\beta_{l,j,K}}(\rho)
=
V_{\beta_{l,j,K}}\rho V_{\beta_{l,j,K}}^\dagger$
does not represent an additional gate inserted into the circuit. Instead, it describes the residual phase channel that remains after the adaptive feed-forward phase has cancelled the known tail phase. In contrast, the factor $Z^{k_{l-1}}$ is the real diagonal operation associated with the ideal value of the current output bit. It induces the free channel
$\mathcal{Z}_{k_{l-1}}(\rho)
=
Z^{k_{l-1}}\rho Z^{k_{l-1}} $.

Thus, the only nontrivial residual phase contribution is contained in
$\Theta_{\beta_{l,j,K}}$.
Accordingly, the original branch-wise sequence
\[
\mathcal{H}
\longrightarrow
\mathcal{U}_{l,j}
\longrightarrow
\mathcal{R}_{l,h}
\longrightarrow
\mathcal{H}
\longrightarrow
\mathcal{M}_{Z}
\]
can be equivalently represented, after fixing $j$, $h$ and $K$, as
\[
\mathcal{H}
\longrightarrow
\mathcal{Z}_{k_{l-1}}
\longrightarrow
\Theta_{\beta_{l,j,K}}
\longrightarrow
\mathcal{H}
\longrightarrow
\mathcal{M}_{Z}.
\]
These two sequences are two equivalent descriptions of the same branch-wise process. They should not be interpreted as two channel chains to be concatenated.
The corresponding branch-wise effective channel scheme is shown in Fig.~2.
For fixed \(j\), \(h=(k_0,\ldots,k_{l-2})\) and \(K\), the controlled modular
operation together with the adaptive feed-forward correction is effectively decomposed
into a real diagonal factor and a residual phase channel.
\begin{figure}[ht]
\centering {\begin{minipage}[figure2]{0.9\linewidth}
\includegraphics[width=1.0\textwidth]{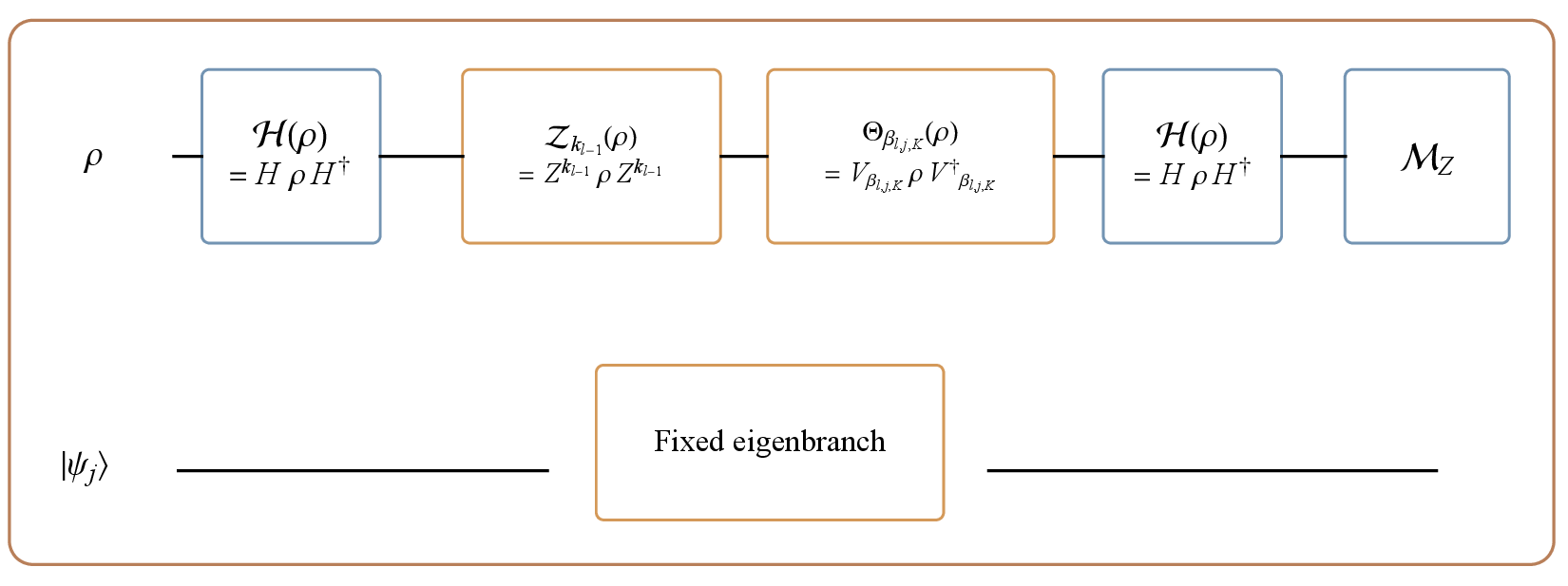}
\end{minipage}}
\caption{
Effective branch-wise channel decomposition of one sequential block.
For fixed \(j\), \(h=(k_0,\ldots,k_{l-2})\) and \(K\), the adaptive correction leaves
the residual phase channel \(\Theta_{\beta_{l,j,K}}\), together with the real diagonal
factor \(\mathcal Z_{k_{l-1}}\).
}
\end{figure}

\noindent{\bf Proposition 1}
Starting with the control qubit in $|0\rangle$, after the initial Hadamard gate, the controlled unitary $CU_l$, and the correct feed-forward phase $R'_l(K)$,
but before the final Hadamard gate, the control qubit state is
\begin{equation}\label{p1}
|\psi_{l,j,K}\rangle
=
\frac{
|0\rangle
+
(-1)^{k_{l-1}}e^{\mathrm{i}\beta_{l,j,K}}|1\rangle
}{\sqrt 2}.
\end{equation}

\noindent{\it Proof.}
The phase induced by $CU_l$ on the eigenbranch $j$ is
\begin{equation*}
e^{\mathrm{i}\theta_{l,j}}
=
\exp\left(
2\pi \mathrm{i}2^{L-l}\frac{j}{r}
\right).
\end{equation*}
By the proof of Theorem 1, we have $e^{\mathrm{i}\theta_{l,j}}=e^{\mathrm{i}2\pi M_l}
e^{\mathrm{i}\pi k_{l-1}}
e^{\mathrm{i}2\pi\phi_l'(K)}
e^{\mathrm{i}\beta_{l,j,K}}$.
Since $e^{\mathrm{i}2\pi M_l}=1$ and $e^{\mathrm{i}\pi k_{l-1}}=(-1)^{k_{l-1}},$
the feed-forward gate $R_l'(K)$ cancels the factor $e^{\mathrm{i}2\pi\phi_l'(K)}$. Thus the remaining control state is (\ref{p1}).
\qed

\noindent{\bf Theorem 2}
The probability that the $l$-th block correctly reads out the bit $k_{l-1}$ is
\begin{equation}\label{t2}
p_{l,j,K}
=
\frac{1+\cos\beta_{l,j,K}}{2}.
\end{equation}

\noindent{\it Proof.}
If $k_{l-1}=0$, then
\begin{equation*}
|\psi_{l,j,K}\rangle
=
\frac{|0\rangle+e^{\mathrm{i}\beta_{l,j,K}}|1\rangle}{\sqrt 2}.
\end{equation*}
After applying the final Hadamard gate,
\begin{equation*}
H|\psi_{l,j,K}\rangle
=
\frac{
(1+e^{\mathrm{i}\beta_{l,j,K}})|0\rangle
+
(1-e^{\mathrm{i}\beta_{l,j,K}})|1\rangle
}{2}.
\end{equation*}
Thus the probability of measuring $0$ is
\begin{equation*}
\left|
\frac{1+e^{\mathrm{i}\beta_{l,j,K}}}{2}
\right|^2
=
\frac{1+\cos\beta_{l,j,K}}{2}.
\end{equation*}
If $k_{l-1}=1$, the correct outcome is $1$, and the same calculation gives the same probability. Hence the probability
is given by (\ref{t2}). \qed

\noindent{\bf Proposition 2}
For a fixed eigenbranch $j$ and a fixed output path
$K\in\{0,1,\ldots,q-1\}$, define
$\delta_{j,K}=\frac{j}{r}-\frac{K}{q}$ with $q=2^L .$
The ideal probability of obtaining the sequential output path $K$,
conditioned on the eigenbranch $j$, is
\begin{equation}\label{p2}
P(K|j)
=
\prod_{l=1}^{L} p_{l,j,K}
=
\prod_{l=1}^{L}
\frac{1+\cos\beta_{l,j,K}}{2}
=
\left[
\frac{\sin(\pi q\delta_{j,K})}
     {q\sin(\pi \delta_{j,K})}
\right]^2,
\end{equation}
where the value is understood as the limiting value $P(K|j)=1$ when
$\delta_{j,K}=0$.

\noindent{\it Proof.}
By Theorem 2, we have
\begin{equation*}
p_{l,j,K}
=
\frac{1+\cos\beta_{l,j,K}}{2}
=
\cos^2\left(\frac{\beta_{l,j,K}}{2}\right).
\end{equation*}
Since
$\beta_{l,j,K}
=
2\pi 2^{L-l}\delta_{j,K},$
we have
\begin{equation*}
P(K|j)
=
\prod_{l=1}^{L}
\cos^2\left(\pi 2^{L-l}\delta_{j,K}\right).
\end{equation*}
Setting $s=L-l$ and using the trigonometric product identity,
\begin{equation*}
\prod_{s=0}^{L-1}\cos(2^s x)
=
\frac{\sin(2^L x)}{2^L\sin x}
\end{equation*}
with $x=\pi\delta_{j,K}$, we obtain (\ref{p2}). \qed

Let $S_r$ denote the set of measurement outcomes $K$ for which the classical
continued-fraction post-processing returns the correct order $r$. Since the
initial work-register state decomposes uniformly over the $r$ eigenbranches,
the ideal success probability is
\begin{equation}
P_{\mathrm{succ}}^{\mathrm{ideal}}
=
\frac{1}{r}
\sum_{j=0}^{r-1}
\sum_{K\in S_r}
P(K|j).
\end{equation}
Using Proposition 2, this becomes
\begin{equation}
P_{\mathrm{succ}}^{\mathrm{ideal}}
=
\frac{1}{r}
\sum_{j=0}^{r-1}
\sum_{K\in S_r}
\left[
\frac{\sin(\pi q\delta_{j,K})}
     {q\sin(\pi \delta_{j,K})}
\right]^2 ,
\end{equation}
with the limiting value $1$ when $\delta_{j,K}=0$.

\vskip0.2in

\noindent {\bf 4. Imaginarity in the full success probability}

In this section, we compute the imaginarity of the residual
phase channel. We then combine this single-block quantity with the branch-wise
probability formula derived in the previous section to obtain the exact ideal
success probability of the sequential order-finding protocol and corresponding
lower and upper bounds.

\noindent{\bf Theorem 3} Let $\Theta_\beta(\rho)=V_\beta\rho
V_\beta^\dagger$, where
$V_\beta=\operatorname{diag}(1,e^{\mathrm{i}\beta})$. Then
\begin{equation}\label{t3}
M_c^t(\Theta_\beta)
=
M_d^t(\Theta_\beta)
=
M_{dc}^t(\Theta_\beta)
=
|\sin\beta|.
\end{equation}

\noindent{\it Proof.}
Write a qubit state $\rho$ in terms of Pauli matrices,
\begin{equation*}
\rho
=\frac{1}{2}(I+xX+yY+zZ),
~~x^2+y^2+z^2\le 1.
\end{equation*}
The phase channel $\Theta_\beta$ rotates the $X$-$Y$ plane,
\begin{equation*}
\Theta_\beta(\rho)
=
\frac{1}{2}
\left[
I
+
(x\cos\beta-y\sin\beta)X
+
(x\sin\beta+y\cos\beta)Y
+
zZ
\right].
\end{equation*}
First consider $M_c^t$. For a real input
\begin{equation*}
\sigma=\Delta(\rho)=\frac{1}{2}(I+xX+zZ),
\end{equation*}
we have
\begin{equation*}
\Theta_\beta(\sigma)=\frac{1}{2}(I+x\cos\beta X+x\sin\beta Y+zZ).
\end{equation*}
Define the free channel
\begin{equation*}
\Phi_\beta(\rho)
=
p\rho+(1-p)Z\rho Z,
~~
p=\frac{1+\cos\beta}{2}.
\end{equation*}
For the same real input, one gets
\begin{equation*}
\Phi_\beta(\sigma) = \frac{1}{2} (I+x\cos\beta X+zZ).
\end{equation*}
Therefore,
\begin{equation*}
(\Theta_\beta\Delta-\Phi_\beta\Delta)(\rho)
=
\frac{x\sin\beta}{2}Y.
\end{equation*}
Since $\|Y\|_1=2$, we get
$\left\|
(\Theta_\beta\Delta-\Phi_\beta\Delta)(\rho)
\right\|_1
=
|x\sin\beta|
\le
|\sin\beta|$.
Thus, $M_c^t(\Theta_\beta)\le |\sin\beta|$.

On the other hand, we have $\Theta_\beta(|+\rangle\langle+|) =
\frac{1}{2}(I+\cos\beta\,X+\sin\beta\,Y)$. In contrast, since
$|+\rangle\langle+|$ is real, for any free channel $\Phi\in
\mathcal{FC}$, the state $\Phi(|+\rangle\langle+|)$ is real and
hence has no $Y$ component. Therefore, the Bloch-vector difference
between $\Theta_\beta(|+\rangle\langle+|)$ and any free output
$\Phi(|+\rangle\langle+|)$ has a $Y$ component equal to $\sin\beta$,
which implies that
$\left\|
\Theta_\beta(|+\rangle\langle+|)
-
\Phi(|+\rangle\langle+|)
\right\|_1
\geq
|\sin\beta|$.
Hence, $M_c^t(\Theta_\beta)=|\sin\beta|$.

Next consider $M_d^t$. The same free channel $\Phi_\beta$ gives
\begin{equation*}
(\Delta\Theta_\beta-\Delta\Phi_\beta)(\rho)
=
-\frac{y\sin\beta}{2}X,
\end{equation*}
and hence $M_d^t(\Theta_\beta)\le |\sin\beta|$.
For the lower bound, taking
$\rho_{\pm}
=\frac{1}{2}(I\pm Y)$ we obtain
\begin{equation*}
\Delta(\rho_+)=\Delta(\rho_-)=\frac{I}{2}.
\end{equation*}
Since $\Delta\Phi=\Delta\Phi\Delta=\Phi\Delta$ for every free channel $\Phi\in \mathcal{FC}$, we get
\[
\Delta\Phi(\rho_+)=\Delta\Phi\Delta(\rho_+)
=\Delta\Phi(I/2)
=\Delta\Phi\Delta(\rho_-)
=\Delta\Phi(\rho_-).
\]
On the other hand,
\[
\Delta\Theta_\beta(\rho_+)=\frac12(I-\sin\beta\,X),~~
\Delta\Theta_\beta(\rho_-)=\frac12(I+\sin\beta\,X).
\]
The trace-norm distance between these two states is \(2|\sin\beta|\). Therefore,
for any common free output, at least one of the two errors has trace norm at least
\(|\sin\beta|\). Hence, $M_d^t(\Theta_\beta)\ge |\sin\beta|$, which, together with $M_d^t(\Theta_\beta)\le |\sin\beta|$, gives rise to
$M_d^t(\Theta_\beta)=|\sin\beta|.$

Finally, since the trace norm of
\begin{equation*}
(\Delta\Theta_\beta-\Theta_\beta\Delta)(\rho)
=-\frac{\sin\beta}{2}(yX+xY)
\end{equation*}
is $|\sin\beta|\sqrt{x^2+y^2}$, and
the upper bound $|\sin\beta|$ is attained by choosing an equatorial
pure state, namely, a state with $z=0$ and $x^2+y^2=1$ (For example,
$|+\rangle$ corresponding to $x=1$ and $y=0$, or
$|+_y\rangle$ corresponding to $x=0$ and $y=1$), we get $M_{dc}^t(\Theta_\beta)=|\sin\beta|$.
Combining above three identities gives the claim. \qed

For the residual phase in the Shor block, we define
\begin{equation}
m_{l,j,K}
=
M_c^t(\Theta_{\beta_{l,j,K}})
=
M_d^t(\Theta_{\beta_{l,j,K}})
=
M_{dc}^t(\Theta_{\beta_{l,j,K}})
=
|\sin\beta_{l,j,K}|.
\end{equation}
Hereafter, good output paths mean the branch-wise nearest-integer paths $K_j$ to $qj/r$. For $q>r^2$ and $\gcd(j,r)=1$, they lie in $S_r$ and satisfy $|\beta_{l,j,K_j}|\leq \pi/2$ for all $l$.

\noindent{\bf Remark 1} On good output paths, where $|\beta_{l,j,K}|\leq \pi/2$, the single-step success probability becomes
\begin{equation}
p_{l,j,K}
=
\frac{1+\sqrt{1-m_{l,j,K}^2}}{2}.
\end{equation}
Thus, on good output paths, the imaginarity
$m_{l,j,K}=|\sin\beta_{l,j,K}|$ determines the single-step success probability. Since this probability decreases as $m_{l,j,K}$ increases, $m_{l,j,K}$ should be interpreted as the magnitude of the imaginary component of the residual phase channel after the adaptive feed-forward correction,
rather than as a resource that improves the success probability.

\noindent{\bf Theorem 4}
Let \(q=2^L>r^2\). For each \(j\) with \(0\le j<r\) and
\(\gcd(j,r)=1\), let \(K_j\) be an integer nearest to \(qj/r\). Then the ideal
sequential order-finding success probability satisfies the following inequality,
\begin{equation}\label{t4}
P_{\mathrm{succ}}^{\mathrm{ideal}}
\ge
\frac{1}{r}
\sum_{\substack{0\le j<r\\ \gcd(j,r)=1}}
\prod_{l=1}^{L}
\frac{1+\sqrt{1-M_c^t(\Theta_{\beta_{l,j,K_j}})^{2}}}{2}.
\end{equation}

\noindent{\it Proof.}
The initial work-register state $|1\rangle$ decomposes uniformly over the $r$ eigenbranches of $U_B$. Thus each branch $j$ has weight $1/r$. For fixed $j$ and $K$, the probability of the output path $K$ is the product of the conditional single-block probabilities,
\begin{equation*}
P(K|j)=\prod_{l=1}^{L}p_{l,j,K}.
\end{equation*}
Using
$p_{l,j,K}
=
\frac{1+\cos\beta_{l,j,K}}{2},$
and summing over all successful outputs $K\in\mathcal S_r$, we obtain
\begin{equation*}
P_{\mathrm{succ}}^{\mathrm{ideal}}
=
\frac{1}{r}
\sum_{j=0}^{r-1}
\sum_{K\in\mathcal S_r}
\prod_{l=1}^{L}
\frac{1+\cos\beta_{l,j,K}}{2}.
\end{equation*}
Then for $q=2^L>r^2$ and each $j$ with $\gcd(j,r)=1$, we have
\begin{equation*}
\left|
\frac{j}{r}
-
\frac{K_j}{q}
\right|
\le
\frac{1}{2q}
<
\frac{1}{2r^2}.
\end{equation*}

By the standard continued-fraction criterion, $j/r$ and hence $r$ can be recovered from $K_j/q$. For these good paths,
\begin{equation*}
|\beta_{l,j,K_j}|
\le
2\pi 2^{L-l}\frac{1}{2q}
=
\frac{\pi}{2^l}
\le
\frac{\pi}{2}.
\end{equation*}
Therefore,
\begin{equation*}
p_{l,j,K_j}
=\frac{1+\sqrt{1-m_{l,j,K_j}^{2}}}{2}.
\end{equation*}
Keeping only these successful paths gives (\ref{t4}). \qed

Moreover, since
$m_{l,j,K_j}
=
|\sin\beta_{l,j,K_j}|
\le
\sin\frac{\pi}{2^l}$
and the function
$
f(m)=\frac{1+\sqrt{1-m^2}}{2}
$
is decreasing on $[0,1]$, we have
\begin{equation*}
p_{l,j,K_j} \ge \frac{1+\cos(\pi/2^l)}{2}.
\end{equation*}
Thus,
\begin{equation*}
\prod_{l=1}^{L}p_{l,j,K_j} \ge \prod_{l=1}^{L}
\frac{1+\cos(\pi/2^l)}{2}=\prod_{l=1}^{L}
\cos^2\left(\frac{\pi}{2^{l+1}}\right).
\end{equation*}
By Vi\`ete's product formula,
$\prod_{l=1}^{\infty}
\cos\left(\frac{\pi}{2^{l+1}}\right)
=
\frac{2}{\pi},$
the finite product is at least $4/\pi^2$. Since there are $\varphi(r)$ integers $j$ satisfying $0\le j<r$ and $\gcd(j,r)=1$, we obtain
\begin{equation*}
P_{\mathrm{succ}}^{\mathrm{ideal}}
\ge
\frac{4}{\pi^2}\frac{\varphi(r)}{r}.
\end{equation*}

Theorem 4 provides a lower bound for the overall ideal success probability in terms of the imaginarity of operations associated with the residual phase channels. Although the bound is built from the product of the probabilities of all sequential blocks, its conclusion concerns the global success probability of the protocol.
Along the good output paths, smaller residual phase-channel imaginarity leads to a larger single-block success probability, and hence to a larger product-type lower bound.
Thus, the theorem shows that, along the selected good output paths, a larger residual phase channel imaginarity decreases the corresponding contribution to the lower bound on the overall ideal success probability.

\noindent{\bf Theorem 5}
The ideal success probability $P_{\rm succ}^{\rm ideal}$ of the sequential order-finding protocol satisfies the following upper bound,
\begin{equation}\label{eq27}
P_{\rm succ}^{\rm ideal}
\leq
\frac{1}{r}
\sum_{j=0}^{r-1}
\sum_{K\in S_r^+}
\prod_{l=1}^{L}
\frac{
1+s_{l,j,K}\sqrt{
1-\left[M_c^t(\Theta_{\beta_{l,j,K}})\right]^2
}
}{2},
\end{equation}
where $S_r^+$ is an enlarged successful output set,
\[
S_r^+
=
\left\{
K:\exists a,\ 0\le a<r,\ \gcd(a,r)=1,\
\left|
\frac{K}{q}-\frac{a}{r}
\right|
<
\frac{1}{r^2}
\right\}.
\]

\noindent{\it Proof.}
The ideal success probability of the sequential order-finding protocol can be
written as
\begin{equation}\label{pp}
P_{\rm succ}^{\rm ideal}
= \frac{1}{r} \sum_{j=0}^{r-1} \sum_{K\in S_r}
\prod_{l=1}^{L} \frac{1+\cos\beta_{l,j,K}}{2}.
\end{equation}
From the necessary condition for the continued-fraction
post-processing to return the order \(r\), if the outcome \(K\) is
successful, there exists an integer \(a\) satisfying \(0\le a<r\)
and \(\gcd(a,r)=1\) such that
\[
\left| \frac{K}{q}-\frac{a}{r} \right| < \frac{1}{r^2}.
\]
Hence, \(S_r\subseteq S_r^+\). Since each block probability
$\tfrac{1+\cos\beta_{l,j,K}}{2}\ge 0$, replacing $S_r$ in (\ref{pp}) with $S_r^+$
gives rise to
\[
P_{\rm succ}^{\rm ideal}
\le \frac{1}{r} \sum_{j=0}^{r-1} \sum_{K\in S_r^+}
\prod_{l=1}^{L} \frac{1+\cos\beta_{l,j,K}}{2}.
\]

Note that $\cos\beta_{l,j,K} = s_{l,j,K}\sqrt{1-m_{l,j,K}^{2}}$,
where $s_{l,j,K} = \operatorname{sgn}(\cos\beta_{l,j,K})$, and
$M_c^t(\Theta_{\beta_{l,j,K}})=|\sin\beta_{l,j,K}|$. We get
\[
\frac{1+\cos\beta_{l,j,K}}{2} =
\frac{1+s_{l,j,K}\sqrt{1-\left[M_c^t(\Theta_{\beta_{l,j,K}})\right]^2}}{2},
\]
which yields (\ref{eq27}). \qed

\noindent{\bf Remark 2} We claim that the right-hand side of inequality
(\ref{eq27}) is strictly smaller than one. Indeed, for fixed $j$,
the conditional outcome distribution is normalized,
\[
\sum_{K=0}^{q-1}\prod_{l=1}^{L}\frac{1+\cos\beta_{l,j,K}}{2}
=\sum_{K=0}^{q-1}P(K\mid j)=1 .
\]
Consider the branch \(j=0\). Here, \(\delta_{0,K}=-K/q\), and by Proposition~2 we have
\[
P(K\mid 0)=\left(\frac{\sin(\pi K)}{q\sin(\pi K/q)}\right)^2
=\begin{cases} 1, & K=0,\\[2pt] 0, & K=1,\dots,q-1.\end{cases}
\]
Therefore, the \(j=0\) branch is entirely concentrated at \(K=0\). Since
\(\gcd(a,r)=1\) forces \(a\ge 1\), we have
\(\left|0/q-a/r\right|=a/r\ge 1/r>1/r^2\) for all admissible \(a\) (as \(r\ge2\)).
Hence, \(0\notin S_r^+\) and $\sum_{K\in S_r^+}P(K\mid 0)=0 $.
For every other branch \(j=1,\dots,r-1\), the restricted sum is
bounded by the full normalized sum, \(\sum_{K\in S_r^+}P(K\mid
j)\le\sum_{K=0}^{q-1}P(K\mid j)=1\). Hence,
\[
\frac{1}{r}\sum_{j=0}^{r-1}\sum_{K\in S_r^+}P(K\mid j)
\le \frac{1}{r}\Big(0+(r-1)\cdot 1\Big)=\frac{r-1}{r}<1 .
\]
Therefore, we have
\[
P_{\rm succ}^{\rm ideal}
\le \frac{1}{r} \sum_{j=0}^{r-1} \sum_{K\in S_r^+}
\prod_{l=1}^{L}
\frac{1+s_{l,j,K}\sqrt{1-\left[M_c^t(\Theta_{\beta_{l,j,K}})\right]^2}}{2}
\le \frac{r-1}{r}<1.
\]

Theorem 5 gives a nontrivial upper bound (strictly less than 1) by
enlarging the successful output set and controlling each block
probability through the residual phase channel imaginarity. This
result shows that the overall success probability is constrained by
the imaginarity of channels in each sequential block. In particular,
on good output paths the single-block factor
$({1+\sqrt{1-[M_c^t(\Theta_{\beta_{l,j,K}})]^2}})/{2}$ is
monotonically decreasing with the imaginarity $M_c^t$. Therefore, larger
residual phase channel imaginarity corresponds to stronger phase
deviation and hence to a smaller attainable success probability bound.

\vskip0.2in

\noindent {\bf 5. Noisy residual phase channels and success probability bounds}

In the preceding sections, the residual phase channel of the $l$-th block is described by
$\Theta_{\beta_{l,j,K}}$. We now consider a phase-covariant noise model. We assume
that the noise does not mix different eigenbranches.

For the $l$-th block, let $\eta_l\in[0,1]$ be the effective residual visibility and let
$\mu_l$ be a coherent phase bias. Define
$
\theta_{l,j,K}=\beta_{l,j,K}+\mu_l .
$
The noisy residual channel is
$\mathcal N_{l,j,K}=\mathcal D_{\eta_l}\circ \Theta_{\theta_{l,j,K}}$,
where
$\Theta_{\theta}(\rho)=V_{\theta}\rho V_{\theta}^{\dagger}$,
$V_{\theta}=\operatorname{diag}(1,e^{\mathrm{i}\theta})$
and
\begin{equation}
\mathcal D_{\eta}(\rho)=\frac{1+\eta}{2}\rho+\frac{1-\eta}{2}Z\rho Z .
\end{equation}
For qubit state $\rho=\frac12(I+xX+yY+zZ)$, we have
\begin{equation*}
\mathcal N_{l,j,K}:(x,y,z)\mapsto
\left(
\eta_l(x\cos\theta_{l,j,K}-y\sin\theta_{l,j,K}),
\eta_l(x\sin\theta_{l,j,K}+y\cos\theta_{l,j,K}),
z
\right).
\end{equation*}
Thus the ideal residual phase vector $(\cos\beta_{l,j,K},\sin\beta_{l,j,K})$ is replaced by
\begin{equation*}
\left(\eta_l\cos(\beta_{l,j,K}+\mu_l),\eta_l\sin(\beta_{l,j,K}+\mu_l)\right).
\end{equation*}

\medskip
\noindent{\bf Proposition 3}
Let $\mathcal N_{\eta,\theta}=\mathcal D_{\eta}\circ\Theta_{\theta}$ with
$0\leq\eta\leq1$. Then
\begin{equation}
M_c^t(\mathcal N_{\eta,\theta})
=
M_d^t(\mathcal N_{\eta,\theta})
=
M_{dc}^t(\mathcal N_{\eta,\theta})
=
\eta|\sin\theta|.
\end{equation}
Consequently, for the $l$-th block,
$\mathcal I_{l,j,K}
=M_c^t(\mathcal N_{l,j,K})
=\eta_l|\sin(\beta_{l,j,K}+\mu_l)|.$

\noindent{\it Proof.}
The channel $\mathcal N_{\eta,\theta}$ maps the Bloch vector as
\begin{equation*}
(x,y,z)\mapsto
\left(
\eta(x\cos\theta-y\sin\theta),
\eta(x\sin\theta+y\cos\theta),
z
\right).
\end{equation*}
For $M_c^t$, take the free channel
$\Phi_{\eta,\theta}(\rho)=p\rho+(1-p)Z\rho Z$ with $p=({1+\eta\cos\theta})/{2}$.
For a real input $\Delta(\rho)=\frac12(I+xX+zZ)$, one has
\begin{equation*}
(\mathcal N_{\eta,\theta}\Delta-\Phi_{\eta,\theta}\Delta)(\rho)
=
\frac{\eta x\sin\theta}{2}Y.
\end{equation*}
Hence
$M_c^t(\mathcal N_{\eta,\theta})\leq \eta|\sin\theta|$.
The inequality $M_c^t(\mathcal N_{\eta,\theta})\geq \eta|\sin\theta|$
is obtained by taking the real input $|+\rangle\langle+|$, since
no free output contains a $Y$ component. Thus,
$M_c^t(\mathcal N_{\eta,\theta})=\eta|\sin\theta|.$

For $M_d^t$, the same free channel gives
\begin{equation*}
(\Delta\mathcal N_{\eta,\theta}-\Delta\Phi_{\eta,\theta})(\rho)
=
-\frac{\eta y\sin\theta}{2}X,
\end{equation*}
and hence
$
M_d^t(\mathcal N_{\eta,\theta})\leq \eta|\sin\theta|.
$
The lower bound follows by comparing the two inputs
$\rho_{\pm}=\frac12(I\pm Y)$. Since $\Delta(\rho_+)=\Delta(\rho_-)=I/2$,
any free operation gives the same deimaginarized output for these two states, whereas
$\Delta\mathcal N_{\eta,\theta}(\rho_+)$ and
$\Delta\mathcal N_{\eta,\theta}(\rho_-)$ differ by trace distance
$2\eta|\sin\theta|$. Thus
$M_d^t(\mathcal N_{\eta,\theta})=\eta|\sin\theta|.$ Finally,
\begin{equation*}
(\Delta\mathcal N_{\eta,\theta}-\mathcal N_{\eta,\theta}\Delta)(\rho)
=
-\frac{\eta\sin\theta}{2}(yX+xY),
\end{equation*}
whose trace norm satisfies
$
\eta|\sin\theta|\sqrt{x^2+y^2}\leq \eta|\sin\theta|,
$
with the equality attained by any equatorial pure states. Hence
$
M_{dc}^t(\mathcal N_{\eta,\theta})=\eta|\sin\theta|.
$
The proof is complete.
\hfill$\square$

\medskip
\noindent{\bf Proposition 4}
For fixed $j$ and $K$, the noisy probability that the $l$-th block correctly reads out
the bit $k_{l-1}$ is
\begin{equation}\label{p4}
p_{l,j,K}^{\rm noise}
=
\frac{1+\eta_l\cos\theta_{l,j,K}}{2}
=
\frac{1+\eta_l\cos(\beta_{l,j,K}+\mu_l)}{2}.
\end{equation}
Moreover, if $\sigma_{l,j,K}^{\rm
noise}=\operatorname{sgn}(\cos\theta_{l,j,K})$, it holds that
$p_{l,j,K}^{\rm noise} = \frac{ 1+\sigma_{l,j,K}^{\rm noise}
\sqrt{\eta_l^2-\mathcal I_{l,j,K}^2}
}{2}.$

\noindent{\it Proof.} After aligning the correct output bit with the
positive $X$ direction, the control qubit has Bloch vector $(1,0,0)$
before the noisy residual channel. After imposing $\mathcal
N_{l,j,K}$ it becomes
$(\eta_l\cos\theta_{l,j,K},\eta_l\sin\theta_{l,j,K},0)$.
The final Hadamard measurement therefore gives (\ref{p4}). \qed

Under the above noise model, the noisy success probability of the sequential
order-finding protocol is
\begin{equation}
P_{\rm succ}^{\rm noise}
=
\frac1r
\sum_{j=0}^{r-1}
\sum_{K\in S_r}
\prod_{l=1}^{L}
\frac{1+\eta_l\cos\theta_{l,j,K}}{2}.
\end{equation}

\medskip
\noindent{\bf Theorem 6}
Let $q=2^L>r^2$. For each $j$ with $0\leq j<r$ and $\gcd(j,r)=1$, let $K_j$ be an
integer nearest to $qj/r$. We have
\begin{equation}
P_{\rm succ}^{\rm noise}
\geq
\frac1r
\sum_{j\in G_{\mu}}
\prod_{l=1}^{L}
\frac{
1+\sqrt{\eta_l^2-M_c^t(\mathcal N_{l,j,K_j})^2}
}{2},
\end{equation}
where $
G_{\mu}
=
\left\{
j:0\leq j<r,\ \gcd(j,r)=1,\
\cos(\beta_{l,j,K_j}+\mu_l)\geq0\ {\rm for\ all}\ l
\right\}.
$

\noindent{\it Proof.}
Since $q>r^2$, for every $j$ with $\gcd(j,r)=1$, the nearest integer $K_j$ satisfies
the standard continued-fraction recovery condition. Hence, $K_j\in S_r$. From the exact
success-probability formula, we have
\begin{equation*}
P_{\rm succ}^{\rm noise}
=
\frac1r
\sum_{j=0}^{r-1}
\sum_{K\in S_r}
P^{\rm noise}(K|j).
\end{equation*}
Keeping only the successful paths $K_j$ with $j\in G_{\mu}$ gives
\begin{equation*}
P_{\rm succ}^{\rm noise}
\geq
\frac1r
\sum_{j\in G_{\mu}}
P^{\rm noise}(K_j|j).
\end{equation*}
For $j\in G_{\mu}$, we get $\cos(\beta_{l,j,K_j}+\mu_l)\geq0$
for all $l$, and therefore
\begin{equation*}
p_{l,j,K_j}^{\rm noise}
=
\frac{
1+\sqrt{\eta_l^2-M_c^t(\mathcal N_{l,j,K_j})^2}
}{2}.
\end{equation*}
Taking the product over all blocks proves the bound.
\hfill$\square$

\medskip
\noindent{\bf Theorem 7} The success probability of the noisy
sequential order-finding protocol satisfies
\begin{equation}\label{eq34}
P_{\rm succ}^{\rm noise}
\leq
\frac1r
\sum_{j=0}^{r-1}
\sum_{K\in S_r^+}
\prod_{l=1}^{L}
\frac{
1+\sigma_{l,j,K}^{\rm noise}\sqrt{\eta_l^2-M_c^t(\mathcal N_{l,j,K})^2}
}{2}.
\end{equation}

\noindent{\it Proof.}
The continued-fraction success condition implies
$S_r\subseteq S_r^+.$
Thus,
\begin{equation}\label{pp1}
P_{\rm succ}^{\rm noise}
\leq
\frac1r
\sum_{j=0}^{r-1}
\sum_{K\in S_r^+}
\prod_{l=1}^{L}
p_{l,j,K}^{\rm noise}.
\end{equation}
For each block, by using the identity
$\eta_l\cos\theta_{l,j,K}
=\sigma_{l,j,K}^{\rm noise}\sqrt{\eta_l^2-\eta_l^2\sin^2\theta_{l,j,K}}$
and the relation
$M_c^t(\mathcal N_{l,j,K})=\eta_l|\sin\theta_{l,j,K}|$,
the noisy block probability can be rewritten as
\begin{equation*}
p_{l,j,K}^{\rm noise}
=\frac{1+\sigma_{l,j,K}^{\rm noise}\sqrt{\eta_l^2-M_c^t(\mathcal N_{l,j,K})^2}}{2}.
\end{equation*}
Substituting this identity into (\ref{pp1}) yields (\ref{eq34}). \qed

\noindent{\bf Remark 3} The right-hand side of Eq. (\ref{eq34}) is
upper bounded by one. In fact, for each fixed eigenbranch $j$, the
noisy sequential measurement gives a normalized output distribution,
\[
\sum_{K=0}^{q-1}
\prod_{l=1}^{L}
\frac{1+\eta_l\cos\theta_{l,j,K}}{2}
=
\sum_{K=0}^{q-1}P^{\rm noise}(K|j)=1.
\]
Since \(S_r^+\subseteq\{0,1,\ldots,q-1\}\), the restricted sum over \(K\in S_r^+\) cannot exceed
the full normalized sum. Hence,
\[
\frac{1}{r}
\sum_{j=0}^{r-1}
\sum_{K\in S_r^+}
P^{\rm noise}(K|j)
\le
1.
\]
The above inequality is strict if, in addition, there exists at
least one eigenbranch \(j\) for which the noisy outcome distribution
assigns positive probability to some \(K\notin S_r^+\). For example,
this is the case when the noisy block probabilities are strictly
positive for all output paths.

Indeed, when the visibility parameters and the coherent phase biases
satisfy $\eta_l=1$ and $\mu_l=0$ for all $l$, the noisy residual
channel $N_{l,j,K}$ reduces to the ideal residual phase channel
$\Theta_{\beta_{l,j,K}}$, and the noisy block probability and the
corresponding lower and upper bounds all reduce to their ideal
counterparts. Correspondingly, the imaginarity of the noisy residual
channel is given by $\eta_l|\sin(\beta_{l,j,K}+\mu_l)|$, which
reduces to the ideal $|\sin\beta_{l,j,K}|$ when
$\eta_l=1$ and $\mu_l=0$. The difference is that, in the noisy case,
the residual phase deviation is affected not only by
$\beta_{l,j,K}$, but also by the visibility parameter $\eta_l$ and
the coherent phase bias $\mu_l$. If $\mu_l=0$, on good paths
with $\cos\beta_{l,j,K}\geq0$, the noisy single-block success
probability is no larger than the ideal one. However, when a
coherent phase bias is present, there is in general no universal
ordering between the ideal and noisy success probabilities, since
$\mu_l$ may either compensate for or enlarge the residual phase
mismatch. Thus, the noisy bounds not only recover the ideal bounds
in the appropriate limit, but also show how the visibility parameter
$\eta_l$ and the coherent phase bias $\mu_l$ modify the relation
between the residual channel imaginarity and the success probability.

\vskip0.1in

\noindent {\bf 6. Conclusions and discussions}

We have studied the sequential version of Shor's order-finding algorithm
from the perspective of the imaginarity of quantum channels. We have
analyzed the branch-wise action of a single block in the sequential
order-finding protocol. By fixing an eigenbranch of the modular
multiplication unitary and a target output path, we have identified the
residual phase left after the adaptive feed-forward correction. This
leads to an effective decomposition into a real diagonal factor and a
residual phase channel. We have derived the corresponding control-qubit
state and the single-block success probability. Our analysis shows
that the imaginarity of the residual phase channel quantifies the
remaining phase error rather than a resource that directly enhances
the success probability. On good output paths, this residual
imaginarity provides a characterization of the corresponding single-block
success probability.

For the phase channel $\Theta_\beta$, the three trace-norm type
imaginarity measures of channels are all equal to $|\sin\beta|$.
When $\beta=0$ or $\pi$, the imaginarity vanishes, indicating that
the channel has no effective imaginary contribution in this sense.
As $|\sin\beta|$ increases, the channel deviates more strongly from
real operations. By combining the branch-wise probabilities, we
have obtained the exact ideal success probability of the sequential
order-finding protocol, together with a standard lower bound and an
upper bound. We have further extended the analysis to noisy residual
phase channels. For the noisy channel $N_{\eta,\theta}$, the
corresponding trace-norm-type imaginarity measures become
$\eta|\sin\theta|$. Based on this expression, we have obtained the lower
and upper bounds for the noisy success probability. These results
clarify the role of residual imaginarity in sequential phase
estimation and help to distinguish residual phase errors from useful
quantum resources. These findings may shed new light on the
study of imaginarity in quantum algorithms and deepen our
understanding of quantum-resource behavior in algorithmic
information processing.

In Ref.\cite{Ahnefeld2022Shor}, Ahnefeld et al. investigated a sequential version of Shor's order-finding algorithm with a fixed overall structure and showed that the abilities of quantum channels to create and detect coherence provide lower and upper bounds on the success probability. Their work identifies dynamical coherence as a useful resource that plays a determining role in the performance of the algorithm within that fixed framework. More recently, in Ref.\cite{Ahnefeld2026}, Ahnefeld et al. considered a more general phase-estimation setting and established a quantitative relation between the coherence of an input state and the optimal achievable performance of phase estimation, showing that any nonzero amount of coherence can provide an operational advantage. Their framework applies to phase-estimation problems beyond any particular implementation of Shor's algorithm. These works demonstrate that the coherence is an important quantum resource for determining the performance of the corresponding protocols.

In contrast, the present work does not focus on the ability of
channels to create or detect coherence. Instead, we analyze the
sequential order-finding process from the perspective of the
imaginarity of quantum channels, with particular attention to the
residual phase that remains after the adaptive feed-forward
correction. We decompose the effective action of each sequential
block into a real diagonal factor and a residual phase channel, and
prove that the three trace-norm induced imaginarity measures of this
residual phase channel are all equal to $|\sin\beta|$. This result
gives a clear operational interpretation of imaginarity in this
algorithmic setting. On good output paths, the residual
imaginarity characterizes the residual phase deviation after
feed-forward correction, rather than acting simply as a resource
that enhances the success probability. In fact, smaller residual
imaginarity corresponds to a higher probability of correctly reading
out the bit in a single block.

\vskip0.1in

\noindent

\subsubsection*{Declaration of competing interest}
\small {The author(s) declared no potential conflicts of interest with respect to the research, authorship, and/or publication of this article.}

\subsubsection*{Data availability}
\small {No new data were created or analysed in this study.}


\subsubsection*{Acknowledgements}
\small {This work was supported by National Natural Science
Foundation of China (Grant Nos. 12561084, 12161056); Natural Science Foundation of Jiangxi Province (Grant No. 20232ACB211003).}

\end{document}